\documentclass[superscriptaddress, twocolumn, prl, aps]{revtex4-1}

\usepackage{graphicx}
\usepackage{amsmath, amsthm, amssymb, bm}
\usepackage{color}      

\begin{document}

\title{Rethinking the Transient Network Concept in Entangled Polymer Rheology}
\author{Wen-Sheng Xu}
\altaffiliation[Permanent address: ]{State Key Laboratory of Polymer Physics and Chemistry, Changchun Institute of Applied Chemistry, Chinese Academy of Sciences, Changchun, Jilin 130022, China}
\affiliation{Center for Nanophase Materials Sciences, Oak Ridge National Laboratory, Oak Ridge, Tennessee 37831, USA}
\author{Christopher N. Lam}
\affiliation{Center for Nanophase Materials Sciences, Oak Ridge National Laboratory, Oak Ridge, Tennessee 37831, USA}
\author{Jan-Michael Y. Carrillo}
\affiliation{Center for Nanophase Materials Sciences, Oak Ridge National Laboratory, Oak Ridge, Tennessee 37831, USA}
\author{Bobby G. Sumpter}
\affiliation{Center for Nanophase Materials Sciences, Oak Ridge National Laboratory, Oak Ridge, Tennessee 37831, USA}
\affiliation{Computational Sciences and Engineering Division, Oak Ridge National Laboratory, Oak Ridge, Tennessee 37831, USA}
\author{Yangyang Wang}
\email{wangy@ornl.gov}
\affiliation{Center for Nanophase Materials Sciences, Oak Ridge National Laboratory, Oak Ridge, Tennessee 37831, USA}

\begin{abstract}
The classical rheological theories of entangled polymeric liquids are built upon two pillars: Gaussian statistics of entanglement strands and the assumption that the stress arises exclusively from the change of intramolecular configuration entropy. We show that these two hypotheses are not supported by molecular dynamics simulations of polymer melts. Specifically, the segment distribution functions at the entanglement length scale and below deviate considerably from the theoretical predictions, in both the equilibrium and deformed states. Further conformational analysis reveals that the intrachain entropic stress at the entanglement length scale is substantially smaller than the total stress, indicative of a considerable contribution from interchain entropy. Lastly, the relation between entanglement strand entropic stress and macroscopic stress exhibits a bifurcation behavior during deformation and stress relaxation, which cannot be accounted for by the classical theories. 
\end{abstract}

\date{\today}

\maketitle

The tube model of Doi and Edwards \cite{DEbook}, which is the most prominent theory of entangled polymers, attributes the mechanical stress entirely to intrachain ``entropic force" arising from individual entanglement strands obeying Gaussian statistics. This treatment finds its root in the classical transient network theories of Green and Tobolsky \cite{GreenTobolsky}, Lodge \cite{ASLodge}, and Yamamoto \cite{Yamamoto}, which can be traced further back to the kinetic theory of rubber elasticity \cite{Treloar}. The first (H1) of the two key assumptions involved in constructing the stress formula of both the tube and transient network theories is that the distribution $\psi(\bm{r})$ of the end-to-end vectors $\bm{r}$ of the entanglement strands can be described by the Gaussian function:
\begin{equation}\label{eq:1}
\psi(\bm{r})=\left(\frac{\xi_e}{\sqrt{\pi}}\right)^3\exp(-\xi_e^2\bm{r}^2),
\end{equation}
where $\xi_e^2 = 3/(2N_e b^2)$, with $b$ being the Kuhn length and $N_e$ the number of Kuhn steps within an entanglement strand. The second assumption (H2) states that the stress arises solely due to the conformational changes of the entanglement strands, leading to the following formula for the viscoelastic stress tensor $\sigma_{\alpha\beta}$ in the limit of moderate deformation:
\begin{equation}\label{eq:2}
\sigma_{\alpha\beta} =2 k_B T \nu_e \xi_e^2 \left\langle r_{\alpha} r_{\beta} \right\rangle,
\end{equation}
where $\nu_e$ is the number density of entanglement strands. Surprisingly, these two central hypotheses of the classical rheological theory of entangled polymeric liquids, which often appear in standard textbooks \cite{DEbook, BirdBook, LarsonBook}, have not been \textit{explicitly} verified by experiments or simulations. 

\begin{figure*}
\centering
\includegraphics[scale=0.15]{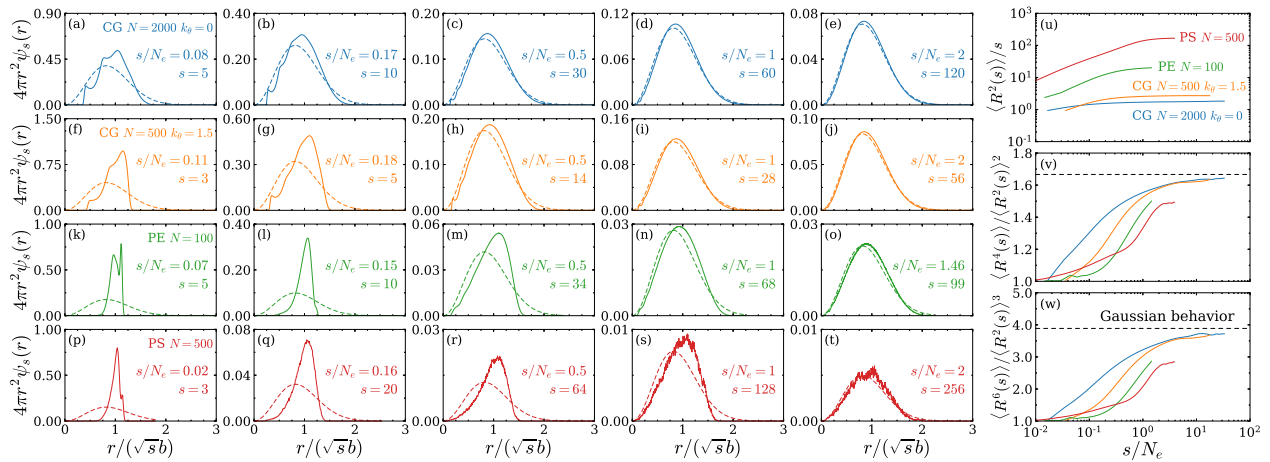}
\caption{(a)-(t) The equilibrium segment distribution $4\pi r^2 \psi_s(r)$ from simulations as a function of $r/(\sqrt{s} b)$ for four different models at various $s$. Solid lines: Simulation data. Dashed lines: Gaussian distributions. (a)-(e) and (d)-(j): CG models of $N=500$ with $k_{\theta}=0$ and $1.5$. (k)-(o): UA-PE model of $N=100$. (p)-(t): CG-PS model of $N=500$. (u)-(w): Analysis of statistical moments of different models as function of $s/N_e$. (u) Normalized mean-squared internal distance $\langle R^2(s) \rangle/s$. (v) and (w) present the ratios $\langle R^4(s)\rangle/\langle R^2(s)\rangle^2$ and $\langle R^6(s)\rangle/\langle R^2(s)\rangle^3$. Horizontal dashed lines: Ideal Gaussian behavior.}
\label{fig:1}
\end{figure*}

Given the fundamental importance of this issue, we have performed a series of equilibrium and non-equilibrium molecular dynamics (MD) simulations on entangled polymer melts using the GPU-accelerated LAMMPS package \cite{Plimpton, Brown, LAMMPS}. To test the first hypothesis, we consider a coarse-grained (CG) bead-spring model of polymer melts, which has been extensively studied by Kremer and Grest \cite{KremerGrest} and many others \cite{HsuKremer}. All beads interact with the WCA potential and the bonded interactions between neighboring beads along the polymer chain are described by the FENE potential. The chain stiffness is controlled by a bending potential $U_{\mathrm{bend}}(\theta)=k_{\theta}(1+\cos\theta)$, where $\theta$ is the angle between two successive bonds. Two values of $k_\theta$, 0 and 1.5, are used in the simulations, resulting in ``fully flexible" and semi-flexible chains, respectively. Polymer melts of $M$ linear chains with $N$ beads are equilibrated at a number density of $\rho = MN/V = 0.85$ and a reduced temperature of $T=1$, similar to the previously reported methods. For $k_{\theta}=0$, two chain lengths were investigated: $M=500$, $N=500$; $M=250$, $N=2000$; for $k_{\theta}=1.5$, $M=500$, $N=500$. We use the estimates, $N_e \approx 60$ \cite{CaoLikhtman} for $k_{\theta}=0$ and $N_e \approx 28$ for $k_{\theta}=1.5$ \cite{HsuKremer, Grest2016}. The major, representative results emerging from these simulations are shown in this Letter and the full details can be found in the Supplemental Material \cite{SM}.

To understand the chain conformation on different length scales, we examine the equilibrium segment distribution function $\psi_s (r)$ at various chemical separations $s=|i-j|$ with $i$ and $j$ being the indices of the beads: $\psi_s(r)=\frac{1}{4\pi r^2 \Delta r}\frac{1}{N-s}\left\langle \sum_{j=1}^{N-s} \delta(r-|\bm{R}_j-\bm{R}_{j+s}|)\right\rangle$. Here, $\bm{R}_j$ is the position vector of bead $j$. For $s = N-1$ with $N \gg N_e$, $\psi_{s=N-1}$, representing the distribution of end-to-end separation of the entire chain, follows the ideal Gaussian distribution. On the other hand, there is plenty of evidence in the literature \cite{HsuKremer, Auhl}, primarily from the mean-squared internal separation, $\langle r^2(s)\rangle_0 =\frac{1}{N-s} \langle \sum_{j=1}^{N-s}(\bm{R}_j-\bm{R}_{j+s})^2\rangle$, that the Gaussian distribution fails at small $s$. For the fully-flexible ($k_{\theta}=0$) and semi-flexible ($k_{\theta}=1.5$) CG bead-spring models, deviation from the ideal Gaussian behavior in $\langle r^2(s)\rangle_0$ becomes apparent when $s \alt 10$. At first glance, these well-known results seem to confirm the Gaussian hypothesis of the network model (H1). However, a detailed analysis of $\psi_s(r)$ paints a different picture. Figure \ref{fig:1} shows the comparison between the $\psi_s(r)$ from simulations (solid lines) and the ideal Gaussian distribution functions (dashed lines) across the entanglement length scale $N_e$. For the Kuhn length $b$ in the Gaussian distribution $\psi_{s,\mathrm{Gaussian}} (\bm{r})=(\xi_s/\sqrt{\pi})^3\exp(-\xi_s^2\bm{r}^2)$, with $\xi_s^2=3/(2sb^2)$, we use $b^2=\langle r^2(s)\rangle_0/s$ as directly determined from the simulations, eliminating any free parameter. It is evident from Fig. \ref{fig:1} that the Gaussian approximation is not fulfilled at the entanglement length scale ($s/N_e=1$) and below ($s/N_e<1$) for these two CG models. In particular, for each $s<N_e$, the distribution function displays a visible oscillatory signature at small $r$, which clearly emanates from the local liquid structure.

To confirm the relevance of these results to real polymer chains, we extend our analysis to a united-atom (UA) model for linear polyethylene (PE) \cite{SKS, Baig2010} as well as a coarse-grained model for polystyrene (PS) \cite{Keten2015, XiaoxiaoJianning}. The PE system consists of 500 chains of $\mathrm{C}_{100}\mathrm{H}_{202}$. The chains are equilibrated at $T=450$ K and $\rho=0.7682$ $\mathrm{g/cm}^3$ with a Nos\'{e}-Hoover thermostat, yielding a pressure of about 1 atm. The equilibrium configurations of the PS melt ($N=500$) are provided to us by the courtesy of Prof. S.-Q. Wang and details of the PS simulation are described elsewhere \cite{XiaoxiaoJianning, XiaoxiaoThesis}. Conformation analysis of these UA-PE and CG-PS chains reveals significant deviation from the Gaussian distribution for $s/N_e \leq 1$ [Fig. \ref{fig:1} (k)-(o) and (p)-(s)]. The Gaussian function is a poor approximation for the distribution function at $s/N_e=1$ and fails completely at $s/N_e=0.5$. This result is in agreement with prior studies of the distribution functions of short chains \cite{MarkCurro1983, MarkCurro1984}. It is worth noting that deviations from the standard Rouse behavior have been observed in neutron spin-echo experiments on an unentangled polyethylene melt and the corresponding atomistic MD simulations \cite{Paul1998, PaulSmith}. These deviations have been attributed to the failure of the \textit{dynamic} Gaussian assumption in calculating the segmental displacement $\bm{R}_i(t)-\bm{R}_j(0)$. Our analysis indicates that the previously observed breakdown of the Rouse model evidently has its origin in \textit{static} structure: non-Gaussian distribution functions at small length scales.

To further quantify the deviation from the Gaussian distribution, we compare several statistical moments of the simulated chains to those of the ideal Gaussian molecules \cite{FloryBook} in Fig. \ref{fig:1}(u)-(w). While the internal separation $\langle R^2(s) \rangle /s$ approaches the ideal Gaussian limit around $N_e$ for all four models, non-Gaussian behavior can be clearly observed at $s \leq N_e$ when higher statistical moments are examined. Figures \ref{fig:1}(v) and (w) show that the deviation from the Gaussian distribution increases with increasing chain stiffness. One might question whether the relatively mild non-Gaussian behavior of the fully-flexible and semi-flexible chain models at $N_e$ bears any rheological significance. However, the deviation from the Gaussian distribution evidently cannot be ignored in the case of the more realistic UA-PE and CG-PS models. Moreover, as we shall demonstrate below, the second assumption (H2) of the transient network picture is not fulfilled in the fully-flexible and semi-flexible chain models. Lastly, we note that our results on the fully-flexible chains are consistent with the classical paper of Kremer and Grest \cite{KremerGrest}, although the authors did not focus on the non-Gaussian behavior and its rheological consequences.   

Both the transient network and tube theories assert that it suffices to coarse-grain entangled polymeric liquids at the entanglement strand level to describe their rheological properties. The underlying assumption is that the deformation of polymer segments at smaller length scales is completely slaved to that of the entanglement strands, given that the time scale associated with the deformation is much longer than the entanglement equilibration time \cite{Watanabe}. This idea is rooted in the classical rubber elasticity theory \cite{James1947} and can be formally expressed using the distribution function of Ullman \cite{Ullman}. For example, for uniaxial extension $\mathbf{E}=\mathrm{diag}(\lambda^{-1/2},\lambda^{-1/2},\lambda)$ applied directly to the entanglement strands, the distribution function $\psi_s(\bm{r})$ for $s/N_e \leq 1$ is:
\begin{eqnarray}\label{eq:4}
\begin{aligned}
\psi_s(\bm{r})= & \left(\frac{\xi_s^2}{\sqrt{\pi}}\right)^3 \prod_{\alpha=x,y,z}\frac{1}{\sqrt{\xi_s^2+\xi_e^2(\lambda_{\alpha}^2-1)}}\\
&\times \exp\left[-\xi_s^4 \sum_{\alpha=x,y,z}\frac{r_{\alpha}^2}{\sqrt{\xi_s^2+\xi_e^2(\lambda_{\alpha}^2-1)}}\right]
\end{aligned}.
\end{eqnarray}
Here, $\lambda$ should be understood as a \textit{microscopic} stretching ratio instead of a macroscopic one. It is straightforward to show that according to Eq. (\ref{eq:4}) the entropic tensile stress at different coarse-graining levels ($s\leq N_e$) is $\Sigma(s) \equiv \sigma_{zz}(s)-\sigma_{xx}(s)=2k_B T \nu_s \xi_s^2 \langle r^2_{z}(s)-r^2_{x}(s)\rangle = \nu_e k_B T(\lambda^2-\lambda^{-1})$, which is equal to the entropic stress of the entanglement strands and independent of $s$. From this discussion, it becomes apparent that in addition to the analysis of equilibrium distribution function $\psi_s(r)$, two other crucial tests can be performed to critically examine the classical transient network concept, and in particular, hypothesis H2. First, using non-equilibrium MD simulation, one can analyze the distribution function $\psi_s(\bm{r})$ at $s/N_e < 1$ to see whether or not it is indeed coupled to the deformation of entanglement strands according to Eq. (\ref{eq:4}). Second, one can examine the entropic stress at different $s$. For $1 \ll s <N_e$, $\sigma_{zz}(s)-\sigma_{xx}(s)$ should be equal to the entropic stress at $s=N_e$ as well as the total stress evaluated directly from the virial formula.

\begin{figure}
\centering
\includegraphics[scale=0.21]{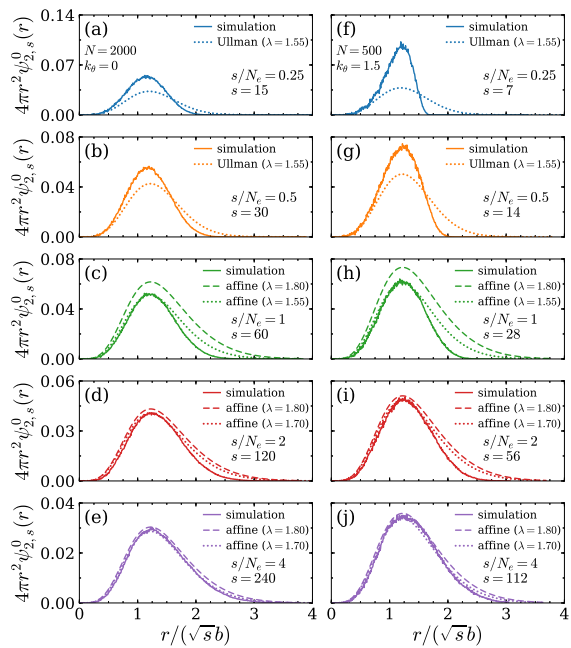}
\caption{$4\pi r^2 \psi_{2,s}^0(r)$ as a function of $r/(\sqrt{s} b)$ at various $s$ for uniaxial extension of $\lambda=1.8$. (a)-(e): Results for $N=2000$ and $k_{\theta}=0$. (f)-(j): Results for $N=500$ and $k_{\theta}=1.5$.}
\label{fig:2}
\end{figure}

To test these ideas, we perform non-equilibrium MD simulations on the two CG models to investigate their uniaxial extension behavior up to stretching ratio $\lambda=3.0$ and the subsequent stress relaxation \cite{SM}, using the deformation protocol described previously \cite{Xu2018}. The equilibrated polymer melt is uniaxially elongated in the $z$-direction with a constant engineering strain rate and the equilibrium pressure of the melt is imposed in the $x$- and $y$-directions via a Nos\'{e}-Hoover barostat. Concomitant with stretching of the box in the tensile direction, the sample shrinks in the perpendicular directions. We employ the spherical harmonic expansion technique \cite{Huang2017, WangPRX, Huang2018} to quantitatively analyze the anisotropic segment distribution function $\psi_s(\bm{r})$: $\psi_s(\bm{r})=\sum_{l,m}\psi_{l,s}^m(r)Y_l^m(\theta,\phi)$, where $\psi_{l,s}^m(r)$ is the expansion coefficient corresponding to each real spherical harmonic function $Y_l^m(\theta,\phi)$. In the case of uniaxial extension, $m=0$ and $l$ is even [28]. Figure \ref{fig:2} shows the leading anisotropic distribution function $4\pi r^2 \psi_{2,s}^0(r)$ of $k_{\theta}=0$, $N=2000$ and $k_{\theta}=1.5$, $N=500$, after a uniaxial extension of $\lambda=1.8$, performed at initial Rouse Weissenberg numbers $\mathrm{Wi}_{R,0}=41.8$ and $5.15$, respectively. To put these results in perspective, we compare the $4\pi r^2 \psi_{2,s}^0(r)$ at $s/N_e \geq 1.0$ from the simulations with those predicted by the deformed Gaussian distribution function \cite{LarsonBook}:
\begin{equation}\label{eq:5}
\psi_s(\bm{r})=\left(\frac{\xi_s}{\sqrt{\pi}}\right)^3\exp\left[-\xi_s^2\left(\lambda r_x^2+\lambda r_y^2 + \frac{r_z^2}{\lambda^2}\right)\right],
\end{equation}
using $\lambda$ as a free fit parameter. At the stretching conditions of these simulations, the molecular deformation at $1 \leq s/N_e \leq 4$ clearly does not follow the macroscopic one (Fig. \ref{fig:2}). For both systems, Eq. (\ref{eq:5}) with a microscopic stretching ratio of $\lambda=1.55$ can provide a reasonable description of the short-distance part of $4\pi r^2 \psi_{2,s}^0(r)$ at the entanglement length scale ($s/N_e=1$) but overestimates the anisotropy at large distances. Accepting that this is the best result we can achieve with Eq. (\ref{eq:5}), we feed this microscopic strain into Eq. (\ref{eq:4}) (Ullman's distribution function) to generate the expected anisotropic distribution functions below the entanglement length scale ($s/N_e<1$). However, $4\pi r^2 \psi_{2,s}^0(r)$ in the simulations deviate strongly from the predicted distributions at these small $s$ (length scales), calling the ideal theoretical picture into question.

Having found no evidence to support the classical coarse-graining hypothesis H2 from the above analysis of the anisotropic distribution functions, we proceed to examine the entropic stress at different $s$. The general expression for the $s$-dependent entropic stress of a Gaussian chain is:
\begin{equation}\label{eq:6}
\sigma_{e,\alpha\beta}(s)=\frac{3\rho k_B T}{s \langle r^2(s)\rangle_0}\left\langle r_{\alpha}(s)r_{\beta}(s)\right\rangle,
\end{equation}
where $\rho$ is the bead (monomer) number density and $\langle \cdots \rangle=\int \cdots \psi_s(\bm{r})d\bm{r}$. Eq. \ref{eq:6} describes how the entropic stress depends on the level of coarse-graining. By examining the statistical average of the dyadic tensor $\bm{rr}$ instead of the detailed form of $\psi_s(\bm{r})$, the entropic stress analysis is a more general test of H2. In the preceding discussions, we show that the equilibrium distribution function $\psi_s(r)$ at $s \leq N_e$ is in fact not Gaussian. Therefore, it is questionable whether Eq. (\ref{eq:6}) can be employed to calculate the intrachain entropic stress at the entanglement length scale and below. Nevertheless, without a better method to accurately determine the entropic stress, we will still use Eq. (\ref{eq:6}) in our stress analysis. Our approach should be understood as a ``proof by contradiction": starting with the assumptions that H1 and H2 are correct, if the results from simulations contradict the theoretical predictions, then the classical transient network concept clearly needs to be reconsidered.

\begin{figure}
\centering
\includegraphics[scale=0.36]{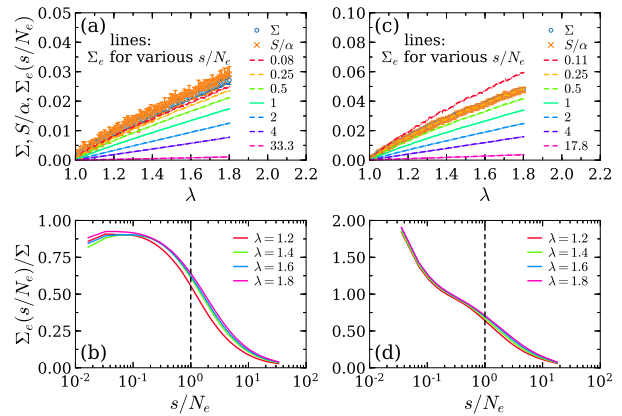}
\caption{(a) Tensile stress $\Sigma$, scaled bond orientation parameter $S/\alpha$, and entropic tensile stress $\Sigma_e(s)$ as a function of stretching ratio $\lambda$ for $N=2000$ and $k_{\theta}=0$. (b) $\Sigma_e(s)/\Sigma$ at various stretching ratios. (c) and (d): Results for $N=500$ and $k_{\theta}=1.5$. The scaling factors $\alpha$ are 0.32 \cite{CaoLikhtman} and 0.75, for the $k_{\theta}=0$ and $k_{\theta}=1.5$ systems, respectively.}
\label{fig:3}
\end{figure}

Figure \ref{fig:3} shows the entropic stress analysis of the uniaxial extension simulations of the fully-flexible and semi-flexible CG models up to $\lambda=1.8$. Because of the moderate strain rates and degrees of deformation, the stress-optical relation is obeyed in these simulations: the tensile stress $\Sigma=\sigma_{zz}-\sigma_{xx}$ and the bond orientation parameter $S$ \cite{CaoLikhtman} are proportional to each other. However, the entropic tensile stress at the entanglement length scale is significantly smaller than the ``macroscopic" stress --- $\Sigma_e(N_e)$ makes up about only 60-70\% of the total stress. Unlike the UA-PE and CG-PS models [Fig. \ref{fig:1}(k)-(t)], the deviation from the Gaussian distribution is relatively small in the two CG models at the entanglement length scale. For this reason, the large discrepancy between $\Sigma_e(N_e)$ and $\Sigma$ is rather surprising. Moreover, contrary to the classical picture, the entropic stress does not level off as $s$ becomes smaller than $N_e$ [Fig. \ref{fig:3}(b) and \ref{fig:3}(d)]. In fact, the $\Sigma_e (s)/\Sigma$ curve does not exhibit any marked changes around $s=N_e$. Lastly, we note that similar behavior is observed during stress relaxation as well \cite{SM}. Our MD simulations of the two CG models clearly do not support the key hypotheses (H1 and H2) associated with the classical transient network picture.

\begin{figure}
\centering
\includegraphics[scale=0.28]{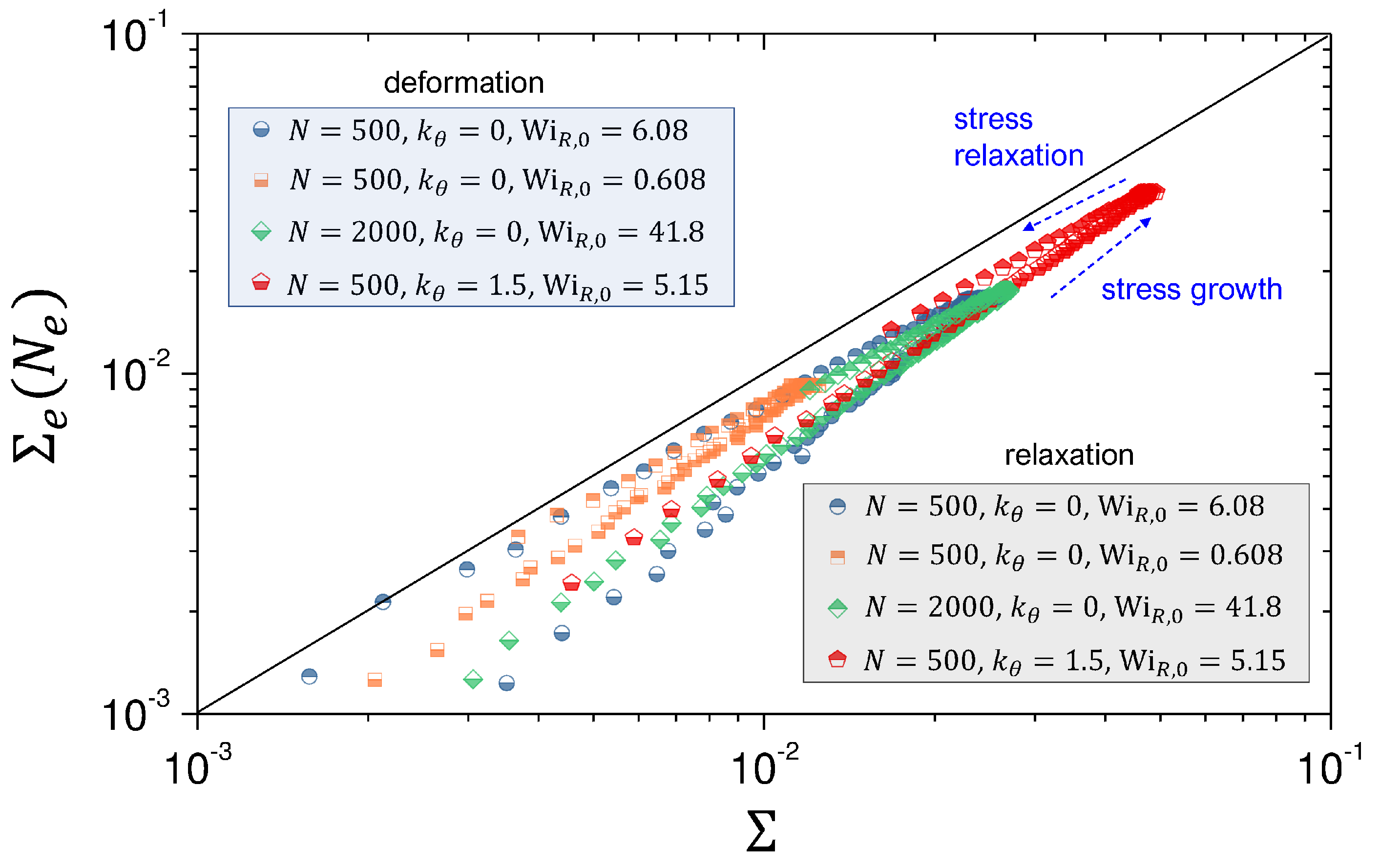}
\caption{Comparison of the entropic tensile stress at the entanglement strand level $\Sigma_e(N_e)$ and the ``macroscopic" stress $\Sigma$ during the deformation (bottom-half-filled symbols) and the subsequent stress relaxation (top-half-filled symbols). The solid line represents the classical transient network picture where $\Sigma_e(N_e) = \Sigma$.}
\label{fig:4}
\end{figure}

For a more comprehensive examination of Eq. \ref{eq:2}, we compare the entropic tensile stress at the entanglement strand level, $\Sigma_e(N_e)$, and the ``macroscopic" stress $\Sigma$ during the deformation and the subsequent stress relaxation in Fig. \ref{fig:4}.  As pointed out previously, the entanglement strand entropic stress is smaller than the total ``macroscopic" stress, when the polymer is not pushed far away from equilibrium \cite{Xu2019}, despite some apparent agreement in steady states \cite{OConnor2018}. This result directly questions the validity of Eq. \ref{eq:2}. Additionally, different relations are found for the entanglement strand entropic stress and the macroscopic stress during deformation and relaxation (Fig. \ref{fig:4}): the deformation and relaxation data reside on two distinct branches. In other words, the correspondence between $\Sigma_e$ and $\Sigma$ is not unique.  This bifurcation behavior is not predicted by any classical theories \cite{DEbook, SM}.

The issue of interchain versus intrachain contributions to stress has long been debated in the literature. MD simulations by Fixman \cite{Fixman}, Gao and Weiner \cite{Gao1, Gao2, Gao3, Gao4}, and Likhtman and coworkers \cite{Likhtman2009, Ramirez2007} have revealed that interchain and excluded volume interactions make significant contributions to stress in dense polymeric liquids, which directly challenges the basic assumption of the tube model. While this finding can in principle be reconciled with the tube model approach by defining an effective bonded force, the analysis of Likhtman \cite{Likhtman2009} shows that even then the cross-correlation of intrachain stresses between different chains is non-negligible. More recently, using a chain model constructed at the level of self-consistently determined primitive paths, Sussman and Schweizer demonstrate that the entanglement plateau modulus can be quantitatively predicted from the correlated intermolecular forces \cite{SussmanSchweizer}. The aforementioned equilibrium MD simulation studies invoke the Green-Kubo relation to evaluate the relaxation modulus $G(t)$. It is therefore not obvious whether the two-time cross-correlation of intrachain stresses between different chains can be interpreted as interchain contributions to stress. In contrast, such complication does not exist in the present non-equilibrium MD simulation method. The ``missing" entropic stress in our analysis (Figs. \ref{fig:3} and \ref{fig:4}) can be viewed as indirect evidence of interchain contributions to stress in entangled polymers.

In summary, this work critically examines two key assumptions of the transient network concept in entangled polymer rheology. Our molecular dynamics simulations show that the conformation distribution function of entanglement strands is non-Gaussian and that the intrachain entropic stress of entanglements does not fully account for the total stress, which challenges the traditional views. While the transient network concept will undoubtedly continue to play an important role in our understanding of entangled polymer rheology, these new results evidently demonstrate the need for a reconsideration of some of the classical theoretical approaches.   

\begin{acknowledgements}
This research (W.-S.X., C.N.L., J.-M.C., and Y.W.) was sponsored by the Laboratory Directed Research and Development Program of Oak Ridge National Laboratory, managed by UT Battelle, LLC, for the U.S. Department of Energy. The data analysis was performed as part of a user proposal at the Center for Nanophase Materials Sciences, which is a DOE Office of Science User Facility. This research used resources of the Oak Ridge Leadership Computing Facility at the Oak Ridge National Laboratory, which is supported by the Office of Science of the U.S. Department of Energy under Contract No. DE-AC05-00OR22725. We thank X. Li, Y. Zheng, and Prof. S.-Q. Wang for sharing with us the equilibrium configurations of PS. We also gratefully acknowledge Prof. K. S. Schweizer for discussions. 
\end{acknowledgements}


\begin{thebibliography}{41}
\bibitem{DEbook} M. Doi and S. F. Edwards, \textit{The Theory of Polymer Dynamics} (Oxford University Press, Oxford, 1986).
\bibitem{GreenTobolsky} M. S. Green and A. V. Tobolsky, J. Chem. Phys. \textbf{14}, 80 (1946).
\bibitem{ASLodge} A. S. Lodge, Trans. Faraday. Soc. \textbf{52}, 120 (1956).
\bibitem{Yamamoto} M. Yamamoto, J. Phys. Soc. Jpn. \textbf{11}, 413 (1956).
\bibitem{Treloar} L. R. G. Treloar, \textit{The Physics of Rubber Elasticity} (Clarendon Press, Oxford, 1975).
\bibitem{BirdBook} R. B. Bird, C. F. Curtiss, R. C. Armstrong, and O. Hassager, \textit{Dynamics of Polymeric Liquids} (John Wiley \& Sons, New York, 1987), 2 edn., Vol. 2, Kinetic Theory.
\bibitem{LarsonBook} R. G. Larson, \textit{Constitutive Equations for Polymer Melts and Solutions} (Butterworth-Heinemann, 2013).
\bibitem{Plimpton} S. Plimpton, J. Comput. Phys. \textbf{117}, 1 (1995).
\bibitem{Brown} W. M. Brown, P. Wang, S. J. Plimpton, and A. N. Tharrington, Comput. Phys. Commun. \textbf{182}, 898 (2011).
\bibitem{LAMMPS}	LAMMPS web page: http://lammps.sandia.gov. The present work utilized the version released on 03/31/2017.
\bibitem{KremerGrest} K. Kremer and G. S. Grest, J. Chem. Phys. \textbf{92}, 5057 (1990).
\bibitem{HsuKremer} H.-P. Hsu and K. Kremer, J. Chem. Phys. \textbf{144}, 154907 (2016).
\bibitem{CaoLikhtman} J. Cao and A. E. Likhtman, ACS Macro Lett. \textbf{4}, 1376 (2015).
\bibitem{Grest2016} G. S. Grest, J. Chem. Phys. \textbf{145}, 141101 (2016).
\bibitem{SM} Supplemental Material.
\bibitem{Auhl} R. Auhl, R. Everaers, G. S. Grest, K. Kremer, and S. J. Plimpton, J. Chem. Phys. \textbf{119}, 12718 (2003).
\bibitem{SKS} J. I. Siepmann, S. Karaborni, and B. Smit, Nature \textbf{365}, 330 (1993).
\bibitem{Baig2010} C. Baig, V. G. Mavrantzas, and M. Kroger, Macromolecules \textbf{43}, 6886 (2010).
\bibitem{Keten2015} D. D. Hsu, W. Xia, S. G. Arturo, and S. Keten, Macromolecules \textbf{48}, 3057 (2015).
\bibitem{XiaoxiaoJianning} X. Li, J. Liu, Z. Liu, M. Tsige, and S.-Q. Wang, Phys. Rev. Lett. \textbf{120}, 077801 (2018).
\bibitem{XiaoxiaoThesis} X. Li, Ph.D. thesis, University of Akron, 2018.
\bibitem{MarkCurro1983} J. E. Mark and J. G. Curro, J. Chem. Phys. \textbf{79}, 5705 (1983).
\bibitem{MarkCurro1984} J. E. Mark and J. G. Curro, J. Chem. Phys. \textbf{81}, 6408 (1984).
\bibitem{Paul1998} W. Paul, G. Smith, D. Y. Yoon, B. Farago, S. Rathgeber, A. Zirkel, L. Willner, and D. Richter, Phys. Rev. Lett. \textbf{80}, 2346 (1998).
\bibitem{PaulSmith} W. Paul and G. D. Smith, Rep. Prog. Phys. \textbf{67}, 1117 (2004).
\bibitem{FloryBook} P. J. Flory, \textit{Statistical Mechanics of Chain Molecules} (Interscience, New York, 1969)
\bibitem{Watanabe} H. Watanabe, Prog. Polym. Sci. \textbf{24}, 1253 (1999).
\bibitem{James1947} H. M. James, J. Chem. Phys. \textbf{15}, 651 (1947).
\bibitem{Ullman} R. Ullman, J. Chem. Phys. \textbf{71}, 436 (1979).
\bibitem{Xu2018} W.-S. Xu, J.-M. Y. Carrillo, C. N. Lam, B. G. Sumpter, and Y. Wang, ACS Macro Lett. \textbf{7}, 190 (2018).
\bibitem{Huang2017} G.-R. Huang et al., Phys. Rev. E \textbf{96}, 022612 (2017).
\bibitem{WangPRX} Z. Wang et al., Phys. Rev. X \textbf{7}, 031003 (2017).
\bibitem{Huang2018} G.-R. Huang, B. Wu, Y. Wang, and W.-R. Chen, Phys. Rev. E \textbf{97}, 012605 (2018).
\bibitem{Fixman} M. Fixman, J. Chem. Phys. \textbf{95}, 1410 (1991).
\bibitem{Gao1} J. Gao and J. Weiner, Macromolecules \textbf{25}, 1348 (1992).
\bibitem{Gao2} J. Gao and J. Weiner, Macromolecules \textbf{27}, 1201 (1994).
\bibitem{Gao3} J. Gao and J. H. Weiner, Science \textbf{266}, 748 (1994).
\bibitem{Gao4} J. Gao and J. H. Weiner, Macromolecules \textbf{29}, 6048 (1996).
\bibitem{Likhtman2009} A. E. Likhtman, J. Non-Newtonian Fluid Mech. \textbf{157}, 158 (2009).
\bibitem{Ramirez2007} J. Ram\'{i}rez, S. K. Sukumaran, and A. E. Likhtman, J. Chem. Phys. \textbf{126}, 244904 (2007).
\bibitem{SussmanSchweizer} D. M. Sussman and K. S. Schweizer, J. Chem. Phys. \textbf{139}, 234904 (2013).
\bibitem{Xu2019} W.-S. Xu, C. N. Lam, J.-M. Y. Carrillo, B. G. Sumpter, and Y. Wang,  Phys. Rev. Lett. \textbf{122}, 059803 (2019).
\bibitem{OConnor2018} T. C. O’Connor, N. J. Alvarez, and M. O. Robbins, Phys. Rev. Lett. \textbf{121}, 047801 (2018).

\end{thebibliography}
\end{document}